# A Magma Accretion Model for the Formation of Oceanic Lithosphere: Implications for Global Heat Loss

Valiya M. Hamza[*], Roberto R. Cardoso[*]. and Carlos H. Alexandrino[**]

[*]Observatório Nacional - MCT, Rio de Janeiro, Brazil

[**]Univ. Federal dos Vales do Jequitinhonha e Mucuri, Teófilo Otoni, Brazil

Corresponding Author: hamza@on.br

**Abstract**

A simple magma accretion model of the oceanic lithosphere is proposed and its implications for understanding the thermal field of oceanic lithosphere examined. The new model (designated Variable Basal Accretion model - VBA) assumes existence of lateral variations in magma accretion rates and temperatures at the boundary zone between the lithosphere and the asthenosphere, similar in character to those observed in magma solidification processes in the upper crust. However, unlike the previous thermal models of the lithosphere, the ratio of advection to conduction heat transfer (the Peclet number) is considered as a space dependent variable. The solution to the problem of variable basal heat input has been obtained by the method of integral transform. The results of VBA model simulations reveal that the thickness of the young lithosphere increases with distance from the ridge axis, at rates faster than those predicted by Half-Space Cooling and Plate models. Another noteworthy feature of the new model is its ability to account for the main observational features in the thermal behavior of both young and old oceanic lithosphere. Thus, heat flow and bathymetry variations calculated on the basis of the VBA model provide vastly improved fits to respective observational datasets. More importantly, the improved fits to bathymetry and heat flow have been achieved for the entire age range and without the need to invoke the ad-hoc hypothesis of large-scale hydrothermal circulation in stable ocean crust. Also, use of VBA model does not lead to artificial discontinuities in the temperature field of the lithosphere, as is the case with GDH (Global Depth Heat Flow) reference models. The results of the VBA model provide a better understanding of the global heat flow variations and estimates of global heat loss. In particular, the model is capable of reproducing regional-scale features in the thermal field of the oceanic crust, identified in recent higher degree spherical harmonic representations of global heat flow. The results suggest that estimates of global heat loss need to be downsized by at least 25%.

## 1. Introduction

Detailed understanding of large-scale variations in the thermal field of the oceanic lithosphere provides important constraints on deep tectonic processes. Nevertheless, thermal models of the lithosphere proposed to date have failed to provide a satisfactory account of some of the important features of large-scale variations in oceanic heat flow. For example, both the Half-Space Cooling [55] and Plate [30] models predict heat flow much higher than the observed values, for young (ages less than 55 Ma) ocean crust. Also, the magnitudes of heat flow anomalies associated with the mid-ocean ridge systems are systematically lower by a factor of 6 at younger ages than those predicted by thermal models proposed in the current literature [41]. In addition, the widths of thermally anomalous zones associated with the spreading centers are narrower (less than 23 Ma) than those calculated (~66 Ma) for a wide range of plausible model parameters. Such discrepancies between model predictions and observational data have given rise to the so-called "oceanic heat flow paradox", for which no satisfactory solution has been found for over the last forty years. The common practice in the current literature is to consider the paradox as originating from eventual perturbing effects of possible regional scale hydrothermal circulation in the ocean crust not accounted for in conventional heat flow measurements (e.g. [41], [54], [56], [27], [60]).

There are however dissenting views on the subject matter of hydrothermal circulation on regional scales [20]. Direct experimental evidences presented to date have confirmed the existence of only isolated pockets of hydrothermal circulation in the central valley and in the rift flanks of spreading centers (e.g. [12], [13], [17], [13], [29], [34]). No direct experimental evidence has so far been presented that point to the existence large scale circulation systems operating in stable ocean crust. Most of the arguments presented to date in favor of the supposed existence of regional-scale convection systems, in parts of ocean floor at large distances from spreading centers, are based on indirect inferences (e.g. [56], [2], [3], [14]).

In the current work, we present a new model of oceanic lithosphere that can overcome the above-mentioned problems and present a satisfactory solution for the heat flow paradox, without the need to invoke the ad-hoc hypothesis of large-scale hydrothermal circulation in stable ocean crust. To place the new model in context, we summarize the main characteristics and inherent limitations of currently accepted thermal models of the lithosphere. Next, the characteristics of thermal fields associated with upwelling of asthenospheric materials are outlined and its compatibility with the new model features examined. Following this, details of the new model fits to observational data on heat flow and bathymetry are presented, along with results of numerical simulations exploring the influence of model parameters. We point out, in addition, that empirical relations such as those proposed for GDH reference models are unnecessary. Implications of the new model results for understanding regional scale variations in global heat flow are discussed and the need to downsize the current estimates of global heat loss emphasized.

## 2. Current Models of the Lithosphere

Thermal models of the lithosphere, with wide acceptance in the current literature, may be classified as falling into essentially two generic groups:

- Half-Space Cooling (HSC) Models; and
- Constant Thickness Plate Models.

In the HSC model the basic assumption is that the temperature of the medium at origin time (t = 0) has a constant value $T_m$ for all depths. This constant $T_m$ then holds for all time at infinite depth. The lithosphere is considered as boundary layer of the mantle convection cells, arising from near surface conductive cooling. The lithosphere (in other words, the boundary layer) grows in thickness continuously as it moves away from the up-welling limb of the mantle convection system. Analytical expressions for temperature variation of the boundary Layer may be obtained as solution to the one-dimensional heat conduction equation [7]. The boundary layer approach has been successful in accounting for first order features in variation of oceanic heat flow with age (e.g. [38], [39], [44]). Nevertheless, this model cannot be considered as satisfactory for several reasons. To begin with the HSC model is strictly valid for heat flux arising from cooling of a stagnant body and not one in which lateral movements occur in response to thermal convection. This is in direct contradiction with one of the essential ingredients of thermal convection, that of lateral movements. In addition, the model predicts infinite heat flow at the ridge axis (the well known problem of singularity in heat flow at time of origin) and heat flow values about five-fold higher than those observed in regions close to the ridge axis. The problem of high model heat flow for young ocean crust is a direct consequence of specific boundary and initial conditions imposed in the Half-Space Cooling model. For example, the assumption of constant temperature for the region beneath the boundary layer imply that asthenosphere resemble an isothermal "magma-ocean". The assumption itself is incompatible with the well-known characteristics of natural convection systems (e.g. [8], [11], [47]). The relatively low heat flow values predicted for older segments of the lithosphere is yet another characteristic feature of the Half-Space Cooling model. It is a consequence of the assumption (e.g. [45], [47]) that boundary layer growth (in other words, the steady advance of the solidification front) takes place exclusively due to heat loss from the upper surface (see discussion in the next section).

It was pointed out by McKenzie [30] that the difficulty with low model heat flow at large distances from the spreading centers can be overcome by assuming that the thickness of the lithosphere at large distances from the ridge axis approaches a constant value. This came to be known later as the *Plate model* and was adopted in several later studies ([32], [39]). In regions close to the ridge axis the interval considered as "Plate" also includes the underlying asthenospheric wedge. The Plate model also assumes that basal temperature is constant beneath vast stretches of stable ocean lithosphere (in other words, asthenosphere is isothermal) and that magma injection occurs only at the central plane of the ridge axis. The Plate model approach has been successful in accounting for the first order features in variation of heat flow with age at large distances from the spreading centers. However, as in the case of the HSC model, it also predicts heat flow values much higher than the observed ones for ocean crust with ages less than 55 Ma.

The assumptions in the Plate model that basal temperature and thickness of the lithosphere are constant rely on the argument that lateral movements of surface layer take place over large nearly isothermal cores present in mantle convection systems. While these may be true of oceanic lithosphere away from spreading centers they can hardly be considered as representative of the thermal structure in regions close to the ridge axis, where non-isothermal conditions are likely to prevail at the base of the lithosphere. In particular, the assumption of constant basal temperature in zones overlying upwelling limbs of asthenosphere contradicts the vast body of observational evidences on temperature variations in intrusive magmatic and thermal metamorphic processes (e.g. [5], [28], [52], [61]). The available experimental data on thermal structures of convective plumes also point to the existence of lateral variations in convective systems (e.g. [35], [36], [59]). Other problems associated with the HSC and Plate models have been discussed in recent works (e.g. [20], [21], [22], [23]).

In an apparent attempt to minimize problems of this type it has been proposed ([39], [53]) that the relations derived from the HSC and Plate models may be combined in such a way that their characteristic constants are compatible with theoretical estimates of heat flow for oceanic crust with ages less than 55 Ma and with experimental heat flow data for ages greater than 55 Ma. In other words, the heat flow – age relations become hybrid in character, a result of the sequential use of solutions of the HSC and Plate models for separate age intervals. The selection of age ranges has been somewhat arbitrary, based on the best match to the data. The hybrid models of Stein and Stein [53] has since then been accepted in the relevant literature as Global Depth Heat Flow (GDH) reference models.

Yet, significant discrepancies continue to exist between the hybrid model values and observational heat flow data, for oceanic lithosphere with ages less than 55 Ma. The current consensus is that such differences arise from perturbing effects of supposed regional scale hydrothermal circulation in the ocean crust, believed to be unaccounted for in conventional heat flow measurements in the oceanic crust ([39], [41], [56], [60]). The origins of some of the problems with the hybrid versions can be traced back to the boundary conditions imposed in the Half-Space Cooling and Plate models. As pointed out recently by Hamza et al [16] GDH reference models imply discontinuities in the deep temperature fields of the lithosphere. In fact the GDH model requires an artificial change in heat flow at 20 My in order to fit the bathymetry data while another artificial change at 55My is necessary to fit the heat flow data [16]. In view of such limitations, the hybrid model approach can hardly be considered as a satisfactory alternative to understanding the thermal structure of the lithosphere.

### 3. Magma Accretion Model of the Oceanic Lithosphere

We consider now a new thermal model of oceanic lithosphere that can overcome some of the shortcomings of the HSC and Plate models discussed in the previous section. Following the premises of the previous models we also assume that lithosphere represents the boundary layer of mantle convection and that its temperatures are always at or below the melting temperature. In developing the new model it is assumed that the growth of this boundary layer, in regions away from the ridge axis, is determined not only by the cooling effects of surface heat loss but also by mass and energy exchange processes taking place at the bottom boundary of the lithosphere. In particular, we consider that the effects of basal magma accretion and lateral temperature variations of the asthenosphere play important roles in the formation of the lithosphere. The new model is designated hereafter as the Variable Basal Accretion model, abbreviated VBA.

The basal accretion may take place as a result of pressure and temperature variations in the ascending magma column, compositional changes occurring during up-flow and differential rates of migration of volatile components. It is well known in fluid dynamics studies ([42], [46]) that the material and thermal exchange processes occurring in the transition zone immediately below the depth of solidification isotherm have variable degrees of viscous coupling with those occurring in regions of fully developed flows. This line of reasoning leads to the deduction that the interaction between the asthenospheric flow and the lower boundary of the lithosphere become less significant as the plate moves away from the ridge axis. Consequently, the amount of heat advected into the basal parts of the lithosphere decrease with distance from the spreading center.

In the present context of developing a new thermal model of the lithosphere the main interest is in examining the effect of basal accretion on the thickness of the lithosphere and on the surface heat flux. If accretion can be considered as a consequence exclusively of conductive heat loss from the upper surface of the lithosphere an approximate description of the boundary layer growth and the ensuing heat flow variation at the surface can be provided on the basis of the well known inverse square root relation of the time elapsed ([7], [37], [38], [49]). This is the classical case of boundary layer growth in stagnant fluids ([8], [9], [11], [51]). However, in cases where accretion is determined also by temperature variations (and ensuing chemical or gravitational differentiation processes) in the stagnant layer (between the solid and liquid parts) the rate of boundary layer growth differs from that in the classical case. Quantitative assessments of such changes in accretion are difficult in view of the limited knowledge of the thermal state of matter and of the chemical and gravitational processes operating at the lithosphere – asthenosphere interface. For the purposes of the present work we assume that the variation in thickness of the lithosphere ($L$) with distance $x$ may be represented by a relation of the type:

$$L(x) = L_0 \left[ 1 - \left( \frac{1}{1 + \eta \sqrt{x}} \right) \right] \tag{1a}$$

where $L_0$ is the stable thickness of the plate at large distances from the ridge axis and $\eta$ an appropriate scaling factor which may be considered as a measure of the change in the degree of basal accretion. Note that at $L(x) = 0$ at $x = 0$ (i.e. at the ridge axis) and $L(x) = L_0$ at $x = \infty$ (i.e. in stable ocean basins). Also, the second member on the RHS of equation (1), given by:

$$L_m(x) = L_0 / \left( 1 + \eta \sqrt{x} \right) \tag{1b}$$

may be considered as the thickness of the column of asthenospheric material between the base of the solid lithosphere and the level of the asthenosphere in stable ocean basins. According to equation (1b) the height of this asthenospheric column decreases from $L_m = L_0$ at $x = 0$ to $L_m = 0$ at $x = \infty$.

The growth of boundary layer is also affected by the presence of lateral temperature variations in the asthenosphere. The assumption of lateral temperature variations is compatible with the vast body of observational evidences on temperature fields in magmatic and thermal metamorphic processes (e.g. [5], [28], [52], [61]) and experimental data on thermal structures of convective plumes ([35], [36], [59]). In discussing geophysical constraints on mantle temperatures Solomon [50] refers to mineral thermometric data for primary magmas of deep origin penetrating the oceanic lithosphere. The conclusion of Solomon [50] is that the asthenospheric temperatures near the spreading centers are in the range 1200 to $1250^0$C, nearly 200 degrees higher than the corresponding values beneath stable ocean basins. Lateral temperature variations are, in fact, a rule rather than an exception in many of the tectonothermal processes in the upper crust. In addition, there are no physically plausible reasons to believe that asthenospheric upwelling take place under isothermal conditions.

In the present case, we assume that the temperature variation in the asthenosphere, along a horizontal plane at the depth corresponding to the base of the stable lithosphere, is best represented by a relation of the type:

$$\left(T_a - T_b\left(x\right)\right)/\left(T_a - T_m\right) = 1 - \exp\left(-C_1\, x\right) \qquad (2)$$

where $T_a$ is the temperature of the asthenosphere in the upwelling regions, $T_b$ (x) its temperature at distance x, $T_m$ its temperature at large distances from the ridge axis and $c_1$ a scaling constant. Note that equation (2) describes the variation of temperature in the upwelling zone of the asthenosphere. The temperature $T_b(x)$ may also be considered as the basal temperature of the interface zone between the asthenosphere and the lithosphere. However, it should not be confused with the temperature at the base of the lithosphere, this latter one remains constant at the solidification temperature $T_S$ and is independent of the distance from the ridge axis.

As mentioned earlier, the main consequence of basal accretion and lateral temperature variations is an increase in the rate of "migration" of the solidification isotherm to larger depths relative to those encountered for isothermal fluids ([9], [51]). As a result the width of the zone of partial melting is narrower in VBA model compared to those in HSC and Plate models. A schematic illustration of this fundamental difference between the VBA and Plate models is illustrated in Figure (1). The VBA model prediction for a narrower width of the magma injection zone seems to be supported experimental heat flow data for oceanic regions. For example, the width of heat flow anomalies in mid-ocean ridge zones usually have dimensions much smaller than those predicted by the Half-Space Cooling (HSC) and Plate models.

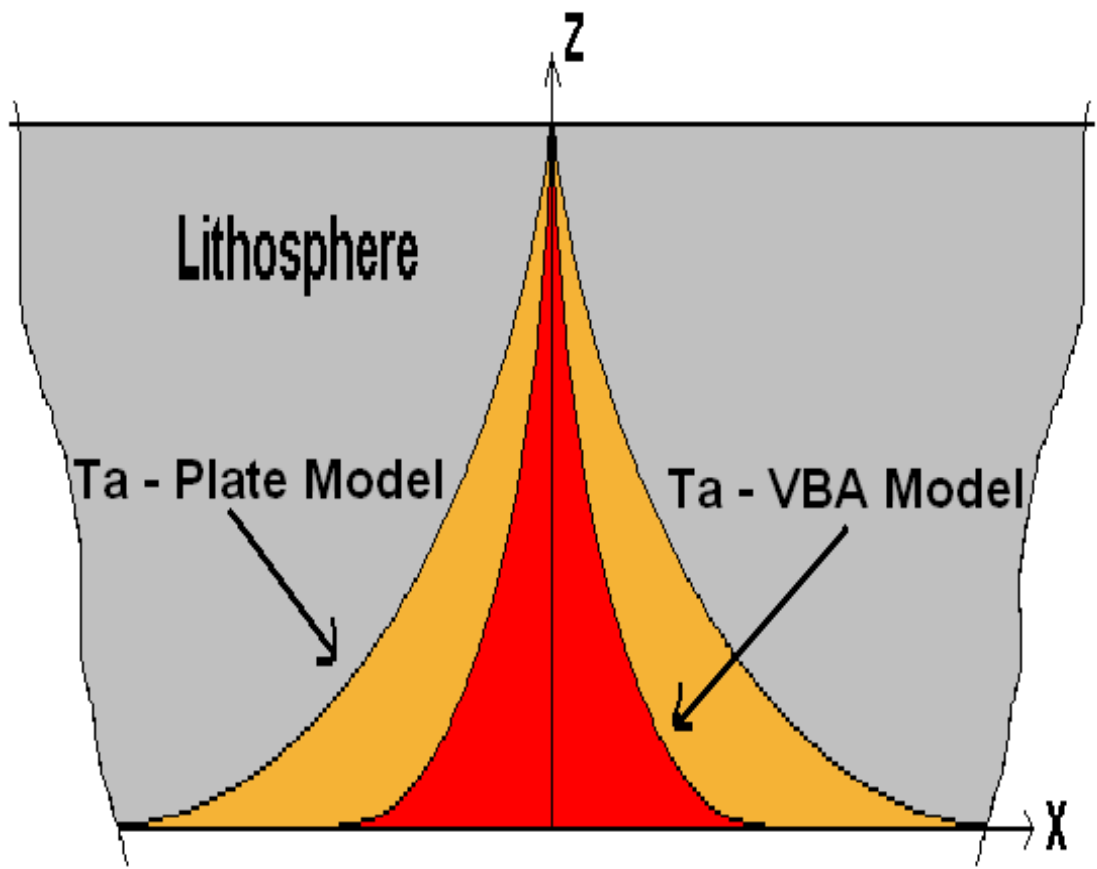

Figure 1: Schematic illustration of solidification isotherms ($T_s$) for cases of constant and variable temperature asthenospheric upwelling. Note that the magma injection zone, whose geometry is determined by the depth to the solidification isotherm, is wider for constant temperature case relative to that for variable temperature.

In the following sections we consider the mathematical basis of the new VBA model and compare the model predictions against observational heat flow and bathymetry data for oceanic regions. In addition, we also compare VBA model values of heat flow and bathymetry with those derived from the half-space cooling and Plate models.

### 3.1. Theoretical Formulation

Consider first the problem of two dimensional heat transfers in a rectangular plate of thickness $L$ moving with velocity $v$ in the horizontal (x) direction. In the model discussed by McKenzie [30] both the thickness of the plate and its basal temperature are assumed to be constants. In the VBA model of the present work these parameters are considered as space dependant variables. As discussed in the previous section, the form of variation of lithosphere thickness ($L$) is assumed to be determined by equation (1) while the systematic decrease in the temperature beneath its base (ie: at the top of the asthenosphere) is assumed to be determined by equation (2). Impositions of these conditions however make the heat transfer problem under consideration non-linear, for which there are no easy analytical solutions. One of the convenient ways of overcoming problems of this type is to make use of the standard method of piece-wise approximation. In this approach, the medium is assumed to be composed of a system of discrete elements, the spatial domains of which are chosen to be sufficiently small that the effects of changes in parameter values (in the present case, the plate thickness and the fusion temperature) may be considered as negligible within each individual element. The relevant differential equation for any particular element of this system is:

$$\rho \, C_p \left[ \frac{\partial T}{\partial t} \; + \; v \; \frac{\partial T}{\partial x} \right] = \lambda \left[ \frac{\partial^2 T}{\partial x^2} + \frac{\partial^2 T}{\partial z^2} \right] + g \begin{array}{l} x_j < x < x_{j+1} \\ t > 0 \\ 0 < z < L \end{array} \tag{3}$$

where $\rho$ is density, $C_P$ the specific heat, $T$ the temperature, $t$ the time, $\lambda$ the thermal conductivity, $g$ the rate of heat generation and $x$ and $z$ the horizontal and vertical coordinates respectively. The origin of the coordinate system is fixed at lower left corner of the rectangular element under consideration. The subscript $j$ refers to the discretization index, which assumes values 0, 1, 2, 3, …n, $n$ being the number of elements. The boundary conditions are:

$$T\left(x, z < 0\right) = T_b\left(x\right) \tag{4a}$$

$$T\left(x, z = 0\right) = T_S \tag{4b}$$

$$T\left(x, z = L\right) = 0 \tag{4c}$$

$$T\left(z, x = 0\right) = T_a \tag{4d}$$

where $T_b(x)$ is the temperature at the top of the asthenosphere, $T_s$ the solidification temperature at the base of the lithosphere and $T_a$ the temperature of the asthenosphere at the ridge axis. Equation (4a) specifies the temperature at the top of the asthenosphere, which is a function of distance from the ridge axis. Equations (4b) and (4c) specify the respective constant temperatures at the base and top of the solid lithosphere. Equation (4d) is the condition that specifies the temperature at the left lateral boundary of the first element. It must be noted that accretion takes place at the solidification temperature $T_s$ so that the base of the lithosphere is determined by the position of the solidification isotherm. Also, for elements situated at large distances $T_b(x) = T_m \approx T_s$, where $T_m$ is the temperature of the asthenosphere in stable ocean basins, at large distances from the ridge axis.

The solution to the problem defined in equations (3) and (4) has been derived using the method of integral transform ([10], [1]). The details of the intermediate steps in the development of the final solution are provided in the Appendix. The relations for temperature ($T$), temperature gradient ($\partial T/\partial z$) and heat flow ($q$), are:

$$T\left(x, z\right) = T_S\left((1 - (z / L)) + \sum_{i=1}^{\infty} \frac{\overline{f}_i \exp\left[a_1\left(x / L\right)\right] \psi_i\left(\mu_i, (z / L)\right)}{N_i\left(\mu_i\right)}\right) \tag{5}$$

$$\frac{\partial T(x, z)}{\partial z} = \frac{T_S}{L}\left[(-1) + \sum_{i=1}^{\infty} \frac{\overline{f}_i \exp\left[a_1\left(x / L\right)\right]}{N_i\left(\mu_i\right)} \frac{d\psi_i}{dz}\right] \tag{6}$$

$$q(x, z) = -\lambda \frac{\partial T(x, z)}{\partial z} = -\lambda \frac{T_S}{L}\left[(-1) + \sum_{i=1}^{\infty} \frac{\overline{f}_i \exp\left[a_1\left(x / L\right)\right]}{N_i\left(\mu_i\right)} \frac{d\psi_i}{dz}\right] \tag{7}$$

where $\psi_i$ are the eigen functions, $\mu_i$ the eigen values and $N$ the norm of the solution. Note that equations (5), (6) and (7) are derived from the corresponding equations (A28), (A29) and (A30) of the dimensionless variables in the Appendix. The terms $f_i$ and $a_1$ are given by the relations (see Appendix):

$$\overline{f_i} = \int_0^1 Z \ \psi_i\left(\mu_i, Z\right) \ dZ \qquad (8)$$

$$a_1 = \left(Pe(x) - \sqrt{Pe^2(x) - 4\mu_i^2}\right)/2 \qquad (9a)$$

In equation (11) $Pe$ is the Peclet number, given by the relation:

$$Pe\left(x\right) = \frac{\rho \ C_p \ v}{\lambda} \ L(x) \qquad\qquad x_j < x < x_{j+1} \qquad (9b)$$

The solutions (5), (6) and (7) are similar to the respective relations derived by McKenzie [30] for the constant temperature Plate model. An important difference is the presence of the coefficient $a_1$ in the exponential terms. The value of this coefficient (see equation 9) depends on the Peclet number, which in turn depends on the plate thickness $L$ (see equations 1 and 10). In implementing the discretization scheme the size of the elements are made sufficiently small, that the Peclet number may be considered as constant within each individual element, allowing thereby use of the solutions (5), (6) and (7). It must however be pointed out that the solutions obtained for the system of discrete elements are coupled in the sense that the solution for element (j+1) is derived from the solution for the previous element (j). The equation relating the temperature profile on the right hand side of the i[th] set of elements constituting the lithospheric block of thickness L (at lateral position X) to the left hand side of the (i+1)[th] set of elements constituting the adjacent block of thickness L+ΔL (at lateral position X+dX) is given by:

$$T(X_j, Z_j)_i = T(X_{j+1}, Z_j)_{i+1} \qquad (10)$$

Transfer of boundary temperature profile from a block of thickness L to the next one with larger thickness L+ΔL leads to a reduction in the value of the temperature gradient. It is a consequence of three parallel processes:

a) The lithospheric block of larger thickness is positioned over a region of asthenosphere with relatively lower temperatures;

b) There is a reduction in the rate of basal accretion of magma; and

c) Loss of heat by the lithospheric block in the vertical direction towards the surface.

Note that the thermal effects of the first and second processes were not taken into account in previous models of the lithosphere (HSC and Plate models), a consequence of the assumption of isothermal asthenosphere in these models.

At this point a brief remark on the scaling constant $\eta$ is in order. Note that the numerical value of this parameter is inversely proportional to the rate of change in basal accretion. Thus, a null value of $\eta$ means that the thickness of the lithosphere is independent of the distance from the ridge axis which implies that the Peclet number is constant. Non-zero positive values of $\eta$ refer to the more general case considered in the present work, where the rates of magma accretion fall off systematically with distance from the ridge axis. Thus, in the VBA model, small values of $\eta$ mean that the total amount of accreted material is large and consequently the approach to stable thermal conditions of the lithosphere is relatively slow. On the other hand, large values of $\eta$ mean higher rates of initial accretion, but relatively rapid approach to the final stable conditions. In such cases, decrease of heat flow with age take place at rates higher than those predicted by the HSC and Plate models. The decrease in thickness of the column of asthenospheric material with distance from the ridge axis (which is inversely proportional to the increase in thickness of the lithosphere), for different values of the basal magma accretion factor $\eta$, is illustrated in Figure (2).

Results of numerical simulations indicate that the VBA model values of heat flow for the case $\eta = 0$ are practically identical to the values derived from the Plate model of McKenzie [30]. This is not altogether surprising since the Plate model assumes that magma accretion occur only at the ridge axis. It is clear that the Plate model envisaged by McKenzie [30] is a particular end-member case of the more general class of VBA models.

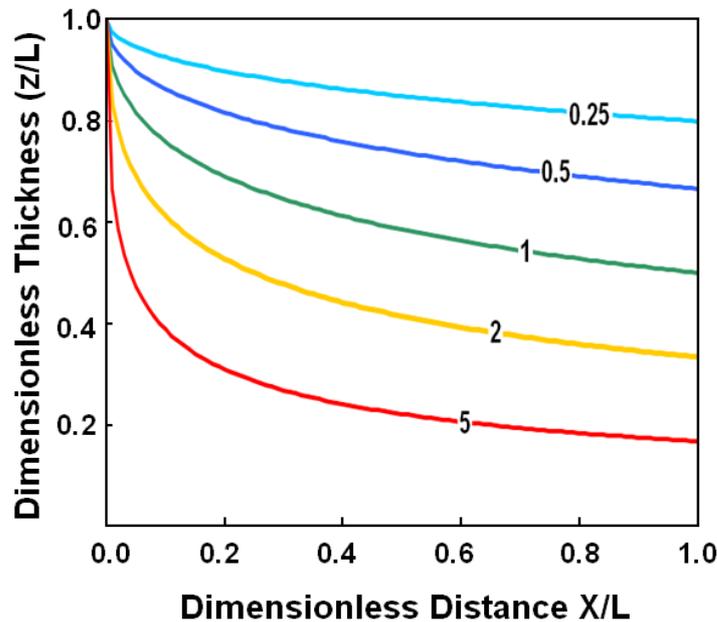

Figure 2: Decrease in thickness of the column of asthenospheric material as a function of the distance from the ridge axis,. The numbers on the curves are values of the parameter (η) in equation (1). See text for details.

The intervals chosen in discretization of VBA model simulations are in the range of 10m to 1km. For gradual changes in the thickness of the lithosphere the computational accuracy of the results obtained in this piece-wise approximation is not overly sensitive to the size of the interval chosen for discretization. On the other hand, the approach has the advantage that the effects of lateral temperature variations arising from compositional changes, which determine the variability in the lower boundary condition (equations 2 and 4a), can be taken into account. A number of numerical simulations were carried out as part of sensitivity tests of model response to parameter values given in Table (1). An example of the results is given in Figure (3), for the case of uniform physical properties of the lithosphere. It reveals that the temperature field is characterized by smooth and continuous changes and devoid of the presence of discontinuities and inversions, provided the changes in temperatures at the top of the asthenosphere are within reasonable limits.

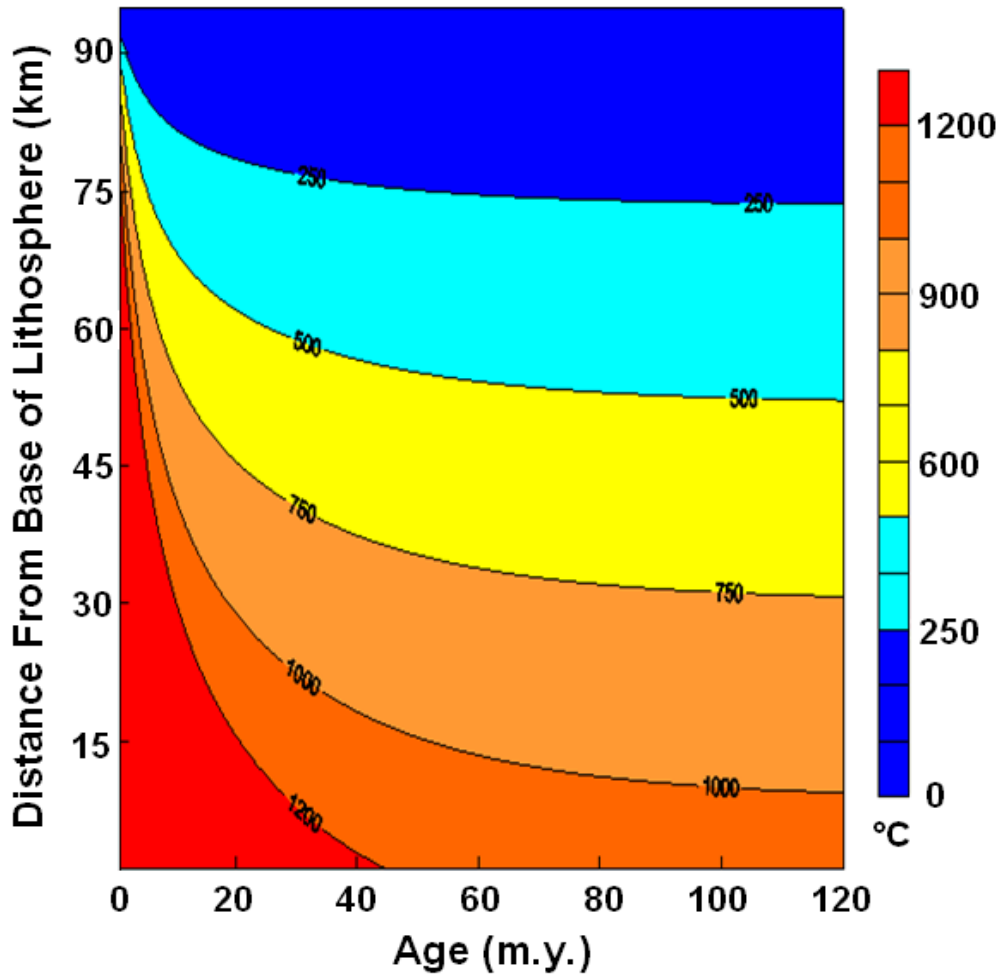

Figure 3: Distribution of isotherms in the oceanic lithosphere, derived from the VBA model of the present work. The numbers on the curves are temperatures in degrees centigrade.

### 3.2. VBA Model Fit to Observational Heat Flow Data

We now make a comparative analysis of the VBA model predictions with results of heat flow measurements in oceanic regions. Following earlier studies (e.g. [53]) we also consider data for oceanic lithosphere with ages less than 160 Ma. The parameter values used in model calculations are given in Table (1), which are essentially identical to those used in earlier studies, allowing thereby direct comparison. Our value of thermal conductivity (λ) is slightly larger, a compromise consistent with results of modern measurements ([40], [19]). The average value of thermal diffusivity ($\kappa$) used (0.8mm$^2$/s in Table - 1) is below the average from 298 to 1300K of modern data for a 60 olivine –30 clinopyroxene – 10% garnet composition ([4], [18]). Our use of lower values than modern averages is consistent with the lower lithosphere exerting a stronger role in controlling heat flow, as it is both more insulating than the upper layers and because heat from the magma must first traverse these deepest lithosphere.

Table 1: Values of parameters used in numerical simulations of the VBA model of the oceanic lithosphere.

| Parameter | Values used in model simulations | |
|---|---|---|
| | Representative | Plausible Range |
| Thickness of Lithosphere | 95 km | 75 – 115 km |
| Thermal Conductivity | 3.3 W m$^{-1}$ °C$^{-1}$ | 3 - 4 W/m/°C |
| Density of Asthenosphere | 3330 kg/m$^3$ | 3300 – 3600 kg/m$^3$ |
| Specific Heat | 1,171 KJ kg$^{-1}$ °C$^{-1}$ | 1000–1500 KJ kg$^{-1}$ °C$^{-1}$ |
| Thermal Diffusivity | 25 x 10$^{-6}$ km$^2$/yr | (11 − 44)x10$^{-6}$ km$^2$/yr |
| Solidification Temperature | 1027 °C | 927 – 1127 °C |
| η (equation 1) | 0.5 km$^{-1/2}$ | 0.4 – 0.7 km$^{-1/2}$ |
| C$_1$ (equation 2) | 0.002 km$^{-1}$ | 0.001 – 0.003 km$^{-1}$ |
| Intervals chosen for discretization of blocks | 10m (Vertical) 1000m (Lateral) | 10 to 1000m |

The variation of VBA model heat flow with age, determined on the basis of equation (7) and the parameter values in Table (1), is illustrated in Figure (4). Also included in this figure are the mean heat flow values reported by Stein and Stein [53] for ocean crust with ages of up to 100 Ma.

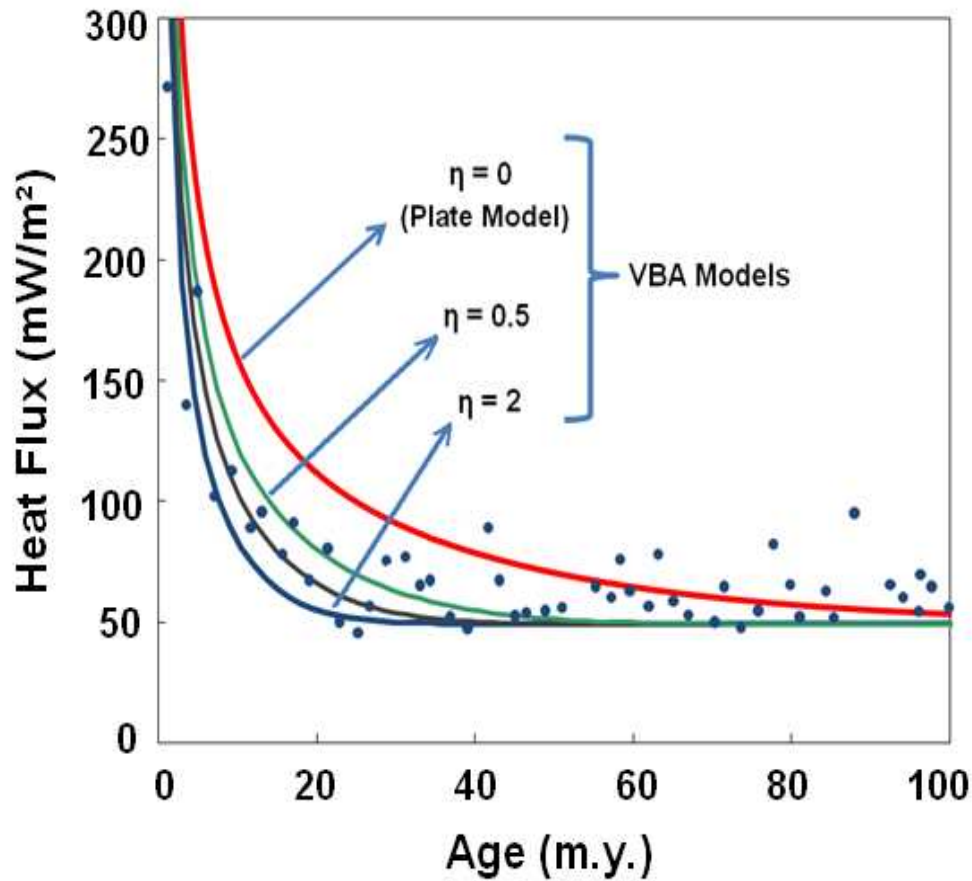

Figure 4: Comparison of VBA model values with experimental heat flow data for the oceanic lithosphere. The model values for basal accretion rates $\eta$ between 0.5 and 2 are indicated by the set of blue, black and green curves. The red colored curve refers to the case $\eta$ =0. It represents the lower end-member case of the family of VBA models, and is identical to that derived from the Plate model [30].

As mentioned earlier, the VBA model leads to a family of solutions depending on the value of the factor $\eta$ that determines the basal accretion rate. The model curves for values of $\eta$ in the range of 0.5 to 2 bracket most of the experimental heat flow data for ocean crust with ages less than 55 Ma. For ages greater than 55 Ma the model curves, independent of the value of $\eta$, tend towards an asymptotic limit. It is clear that VBA Model is capable of providing vastly improved fit to marine heat flow data, relative to that which can be achieved within the framework of HSC and Plate models.

We now examine the dependence of the VBA model predictions on the plausible variations in the values of the main parameters. For this purpose, a number of numerical simulations were carried out for a range of values of the thickness of the plate ($L$), temperature of the stable segment of the asthenosphere ($T_i$) and thermal diffusivity ($\kappa$). Results of some of the numerical simulations are illustrated in the set of panels in figure (5).

Note that changes in thickness of the lithosphere (upper panel of Figure 5) and its basal temperature (middle panel of Figure 5) have only minor influence on the VBA model results. However, the value of thermal diffusivity is found to have a marked influence on the model predictions, as can be seen in the results illustrated in the lower panel of Figure (5). The diffusivity values compatible with experimental heat flow data lie in the range of 11 to $43\times10^{-6}$ km$^2$/yr, which is in reasonable agreement with the values of thermal diffusivity of representative mantle minerals ([40], [22]). The highest and lowest values of $\kappa$ used in our test are extremes observed at $25^0$C and $1027^0$C. This exercise shows that the temperature dependence of $\kappa$; had it been taken into account in a much more complex model, would provide model heat flow values that are consistent with experiments. Reproduction of observational data by the VBA model is not predicated on the specific parameters used in our calculations: values of $L$, $\kappa$; $\lambda$ and $T_i$ that reasonably represent the lithosphere all lead to excellent agreement of the model with the observables.

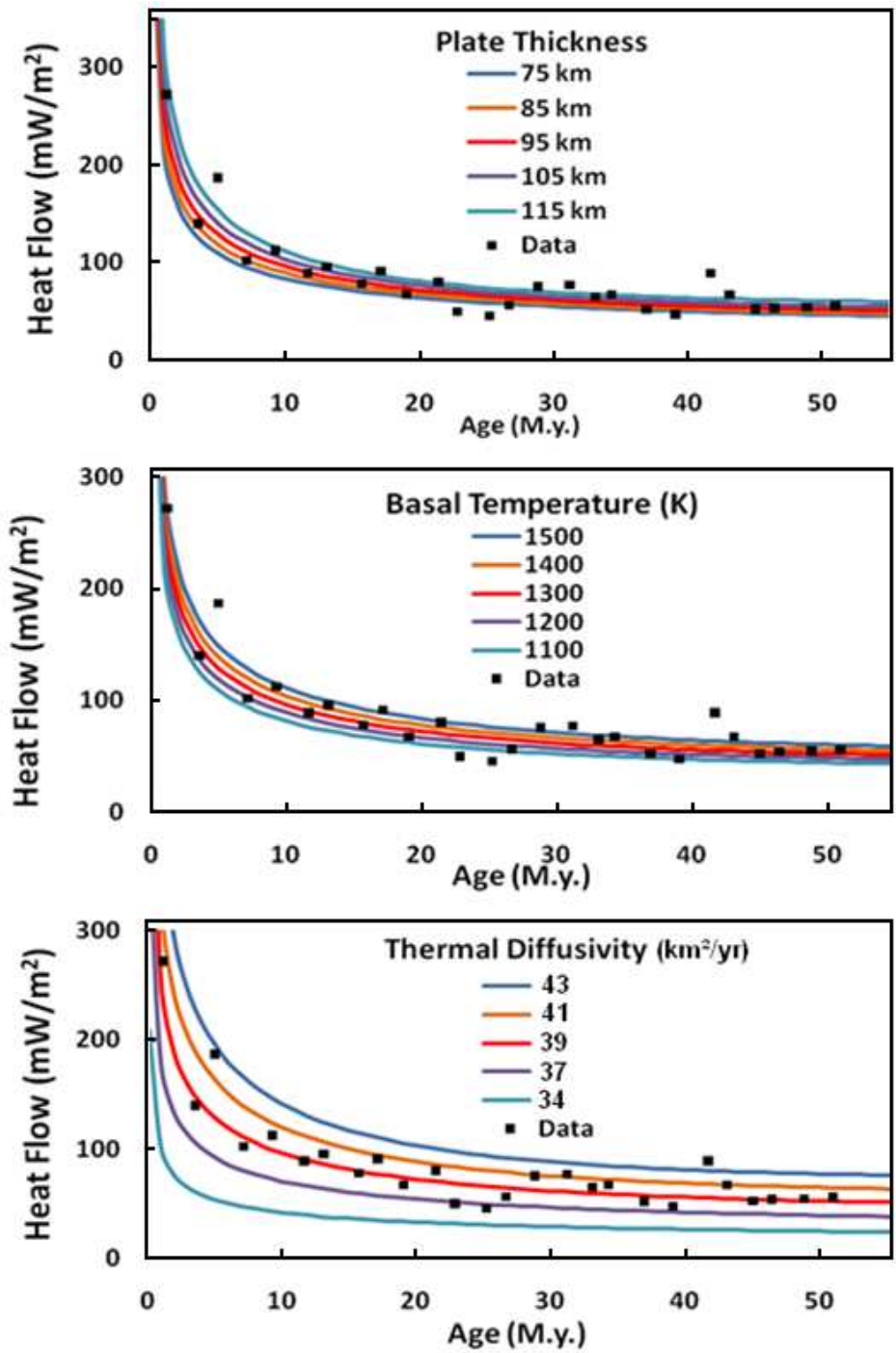

Figure 5: Results of numerical simulations illustrating the dependence of VBA model response to changes in the values assumed for plate thickness (top panel), basal temperatures (middle panel) and thermal diffusivity (lower panel). The dark squares are the mean oceanic heat flow values.

### 3.3. Comparison between Plate and VBA Models

The classical solutions for transient temperature distributions in a plate with constant boundary temperatures have been discussed extensively in the literature ([7], [37]). The Plate model of McKenzie [30] assumes that the transient temperature at the left lateral boundary is equal to the difference between the bottom boundary temperature and the steady state geotherm. The transient temperature in this case is given by the relation (McKenzie, [30]):

$$T_t(z,t) = \frac{2\,T_a}{\pi} \sum_{n=0}^{\infty} \frac{(-1)^n}{n} \sin\left(\frac{n\,\pi\,z}{L}\right) \exp\left(\frac{-k\;n^2\,\pi^2\,t}{L^2}\right) \tag{11}$$

where $T_a$ is the temperature of the lateral boundary, $L$ the plate thickness and $\kappa$ the thermal diffusivity of the medium.

A direct comparison between the transient component of VBA model solution (second term on the RHS of equation-5) and the solution for Plate Model (equation-11) is not straightforward. It is more convenient in this context to adopt the Fourier series solution approach of McKenzie [30] and recast the solution of equation (3) in a form similar to that proposed recently by Cardoso and Hamza [6] for the variable basal heat input problem:

$$T_t(z,t) = \frac{2\,T_a}{\pi} \sum_{n=1}^{\infty} \frac{1}{n} \sin\left(\frac{n\,\pi\,(L-z)}{L\,\eta\,\sqrt{x}}\right) e^{-\frac{\kappa\,n^2\,\pi^2\,t}{L^2}} \tag{12}$$

where η is a scaling constant (with dimensions of $m^{-1/2}$) and $x$ the distance from ridge axis. It is fairly straightforward to show that solution (12) satisfies the initial and boundary conditions of the relevant heat conduction equation of the Plate model. Results of numerical simulations show that the transient temperatures calculated using the solution (12) are nearly identical to the corresponding values obtained from the second term on the RHS of equation (5), provided an appropriate value is chosen for the scaling constant η. In the present case, the value chosen for η is 0.4. The advantage of using equation (12) is that a direct comparison is now possible against the solution obtained for the constant temperature Plate model of McKenzie [30].

There are several important differences between the solutions (11) and (12). To begin with we note that the source strength of the transient component (determined by the pre exponential terms in the summation series of equation (11)) in the Plate model is independent of the distance from ridge axis. Consequently, the magnitude of the transient perturbation in the Plate model remains relatively high for large distances from the ridge axis. On the other hand, the source strength of transient perturbation in VBA model (determined by the pre exponential terms in the summation series of equation (12)) is dependent on the distance from ridge axis, and in addition the argument of the sine function is scaled down by the factor $(L - z) / (\eta \sqrt{x})$. Thus, the magnitude of transient perturbation is relatively small for all times in the VBA model compared to that of the Plate model.

Another important difference between the solutions (11) and (12) may be illustrated by considering the vertical distribution of transient temperature components. As can easily be verified from equation (11) the transient component in the Plate model has a maximum in the upper parts of the lithosphere but is zero at its top and bottom boundaries, a consequence of the imposed constant temperature boundary conditions. Initially, the vertical distribution is asymmetric with respect to the central plane of the lithosphere, as illustrated in Figure (6). The asymmetry is a consequence of the fact that the transient component at the left lateral boundary is set as the difference between the bottom boundary temperature and the steady state geotherm. Nevertheless, the overall shape of the transient response is quite different from that expected for perturbations arising from high temperature intrusions penetrating the base of a relatively cold plate ([7], [24], [25], [37]).

The fact that the transient component is absent for all times at the base of the lithosphere is an indication that the Plate model is inadequate in providing a satisfactory description of heat transfer processes at the lithosphere – asthenosphere boundary, mainly in regions close to the ridge axis. On the other hand, the vertical distribution of the transient component in the VBA model, illustrated in Figure (7), reveal that it is relatively large at the base of the plate and zero at the top boundary. Its overall shape is thus similar to that expected for perturbations generated by a relatively high temperature intrusive penetrating the lower boundary of the lithosphere.

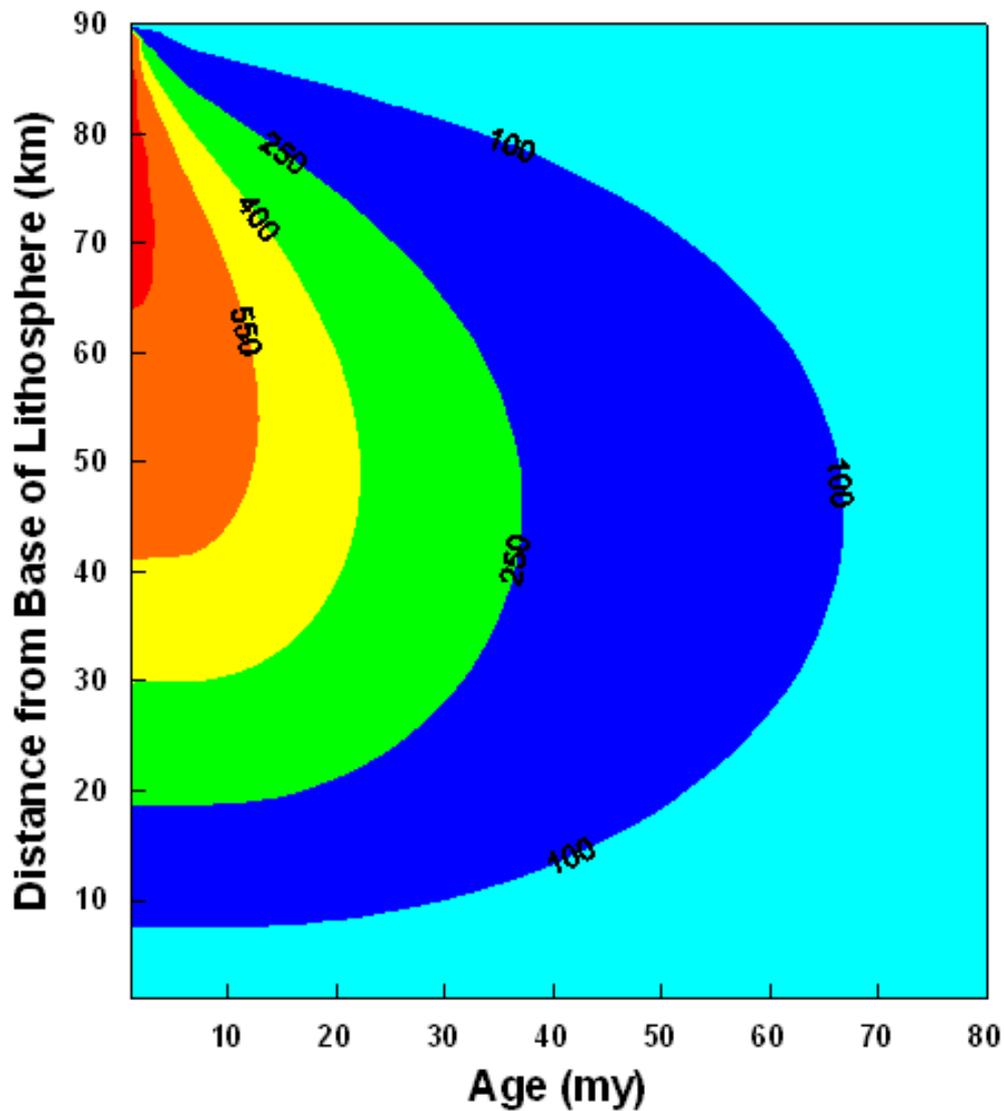

Figure 6: Vertical distribution of transient temperatures in the Plate model (McKenzie, [30]), for the parameter values listed in Table (1). The numbers on the curves are temperatures in degrees centigrade.

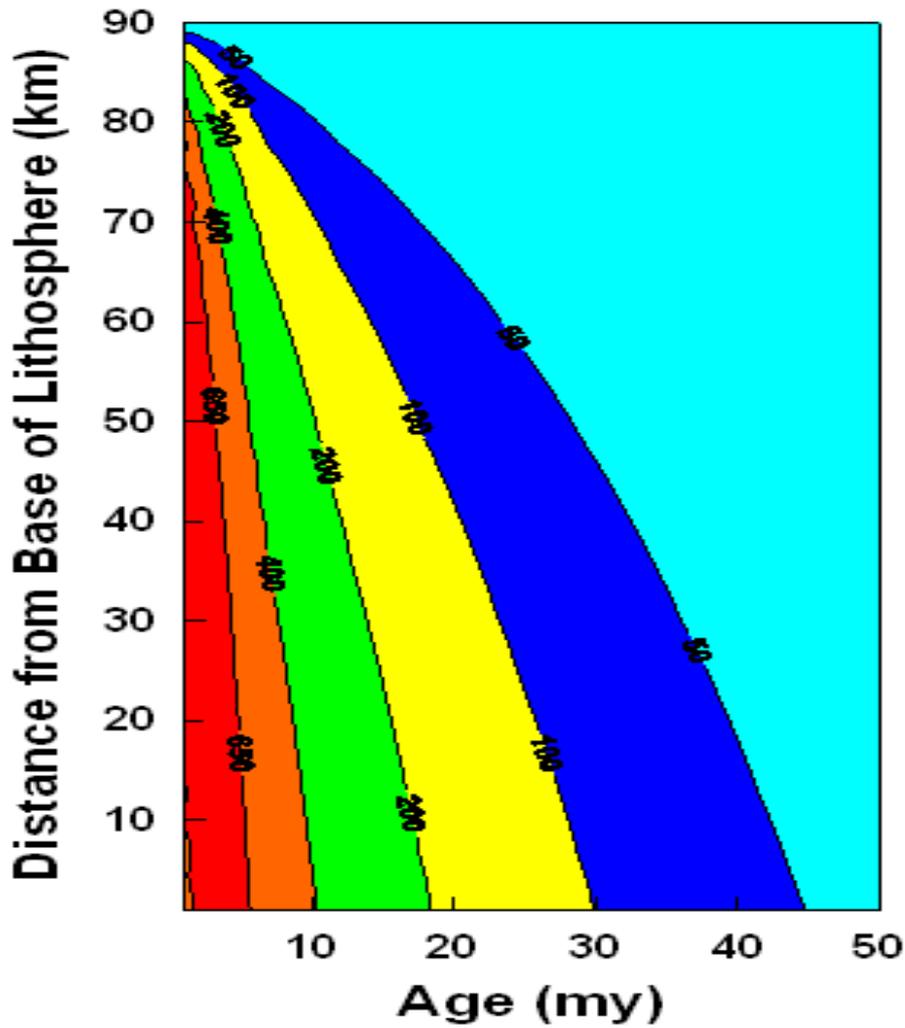

Figure 7: Vertical distribution of transient temperatures in the VBA model of the present work, for the parameter values listed in Table (1). The numbers on the curves are temperatures in degrees centigrade.

Including the steady state component in equation (12) leads to the complete solution for temperature:

$$T(z,t) = T_a\left(1 - \frac{z}{L}\right) + \frac{2 T_a}{\pi} \sum_{n=1}^{\infty} \frac{1}{n} \sin\left(\frac{n \pi (L - z)}{L \, \eta \sqrt{x}}\right) e^{-\frac{\kappa \, n^2 \pi^2 t}{L^2}}$$ (13)

Note that, with the exception of the argument of the sinusoid, equation (13) is similar to the relation derived by Royden and Keen [43] for a lithosphere undergoing stretching due to intrusions. The relation for heat flux, derived from (13) is:

$$q(t) = \lambda \frac{T_a}{L} \left( 1 + \frac{2}{\eta \sqrt{x}} \sum_{n=1}^{\infty} e^{-\frac{\kappa n^2 \pi^2 t}{L^2}} \right)$$

(14a)

where $\lambda$ is the thermal conductivity. For n > 1, the higher order terms in the summation series on the right hand side of equation (14a) are practically negligible. It may therefore be simplified as:

$$q(t) = \lambda \frac{T_a}{L} \left( 1 + \frac{2}{\eta \sqrt{x}} e^{-\frac{\kappa \pi^2 t}{L^2}} \right)$$

(14b)

An interesting aspect of equations (13) and (14) is that these exhibit features similar (though not identical) in character to those of the HSC and Plate models, depending on the time scale chosen. For example, for short times (i.e.: for ages less than 55 Ma), the transient parts of the solutions in equations (13) and (14) are determined mainly by the multiplication factors of the summation, the exponential terms themselves being much less significant. In this case, the temperature and heat flow variations are inversely proportional to the square root of the distance (equivalently, the age of the oceanic crust). Hence the behavior of the thermal field is similar to that predicted by the HSC model. On the other hand, for large times (i.e.: for ages greater than 55 Ma), the transient part of the solution is determined mainly by the exponential terms. In this case the temperature and heat flow variations are similar to those predicted by the Plate model.

Note that the derivation of equations (13) and (14) assumes constant $\lambda$, and hence, $\kappa$ represents the average thermal diffusivity over the lithosphere. Although Equations (13) and (14) pertain to the limit z = L, suggesting that room temperature values for $\lambda$ may be the most appropriate, it can also be argued that average values are consistent with a constant $\kappa$- $\lambda$ derivation. Average values were used in all previous Plate and Boundary Layer models.

**4. VBA Model Fit to Bathymetry Data**

Fits to data for ocean floor bathymetry variations rather than that for surface heat flow is often considered as a relatively more rigorous test of thermal models of the lithosphere. The relation for bathymetry in VBA model has been developed following the isostatic compensation scheme discussed in earlier studies (e.g. [30], [48], [39]). Consider, for example, the mass balance relations for two columns: one situated at the ridge and the other one away from the ridge. For a constant transverse section of the lithosphere the relation for isostatic balance between these columns is:

$$\left(e(x) - d_r\right)\left(\rho_0 - \rho_w\right) = \rho_0\,\beta \int\limits_{\varepsilon(x)}^{e(x)} \Delta T\,dh \tag{15}$$

Where $\rho_w$ is the density of sea water, $d_r$ the elevation of the ocean ridge above the final level to which the lithosphere subsides and $\rho_{ast}$ the density of the asthenosphere. The terms $e(x)$ and $\varepsilon(x)$ represent, respectively, the elevation of the sea floor and the elevation of the base of the lithosphere, $dh$ the thickness of infinitesimal volume element where the temperature change is taking place, $\rho_0$ the reference density of the base of the lithosphere, $\beta$ the volumetric expansion coefficient and $\Delta T$ the temperature difference between the base of the lithosphere and the volume element. Equation (15) has been derived assuming that $\rho_{ast} \approx \rho_0$.

The integration in equation (15) is to be carried out over the entire thickness of the lithosphere whose basal temperature is $T_i$. Developing the integral on the RHS of equation (15), after substitution for the temperature from equation (13), we have:

$$\int\limits_{\varepsilon(x)}^{e(x)} \Delta T\,dh = I_1 - I_2 \tag{16}$$

where

$$I_1 = \int\limits_{\varepsilon(x)}^{e(x)} \left( T_a - \left( T_a\left(1 - \frac{z}{L}\right)\right)\right) dh \tag{17}$$

and

$$I_2 = \int_{\varepsilon(x)}^{e(x)} \left( \frac{2T_a}{\pi} \sum_{n=0}^{\infty} \frac{1}{n} \sin\left( \frac{n\,\pi\,(L-z)}{L\eta\sqrt{x}} \right) e^{-\frac{\kappa n^2 \pi^2 t}{a^2}} \right) dh \qquad (18)$$

The integral $I_1$ in equation (17) represents contraction of the lithosphere in zones without significant magma injection. The limits of the integration are $\varepsilon(x) = 0$ and $e(x) = L$. The result is straightforward:

$$I_1 = \frac{T_a\,L}{2} \qquad (19)$$

On the other hand, the integral $I_2$ in equation (18) represents the transient thermal effect of magma injection and its solution can be obtained by introducing a change of variable that takes into consideration lateral increase in the thickness ($h$) of the solid lithosphere. Since for any specific depth $z$ there are different values of $h$ the relation between $z$ and $h$ may be expressed as:

$$h = L - \frac{(L-z)}{\sigma\sqrt{x}} \qquad (20)$$

where $\sigma$ is the constant of proportionality which has units of m$^{-1/2}$. Substituting (20) in (18), changing the limits of integration and solving for the integral:

$$I_2 = -\frac{2LT_a\,\sigma\eta}{\pi^2} \sum_{n=1}^{\infty} \frac{1}{n^2}\, \cos\left( \frac{n\pi z}{L} \right) \Big|_0^{L+e(x)} e^{-\frac{\kappa n^2 \pi^2 t}{L^2}} \qquad (21a)$$

We designate the product ($\sigma\,\eta$) as the "bathymetry constant" $\delta$ and since $L + e(x) \cong L$, we have

$$I_2 = \frac{4LT_a\,\delta}{\pi^2} \sum_{n=0}^{\infty} \frac{1}{(2n+1)^2}\, e^{-\frac{\kappa(2n+1)^2 \pi^2 t}{L^2}} \qquad (21b)$$

Substituting (19) and (21b) in (15) and after some obvious simplifications we have:

$$\left( e(x) - d_r \right) = \frac{\rho_0\,\beta\,T_a\,L}{2\left(\rho_0 - \rho_w\right)} \left( 1 - \frac{8\,\delta}{\pi^2} \sum_{n=0}^{\infty} \frac{1}{(2n+1)^2}\, e^{-\frac{\kappa(2n+1)^2 \pi^2 t}{L^2}} \right) \qquad (22)$$

Designating $d_s = \frac{T_a\,L\,\rho_0\,\beta}{2\left(\rho_0 - \rho_w\right)}$ we have the final solution:

$$e(x) = \left(d_r + d_s\right) - \frac{8 d_s \, \delta}{\pi^2} \left( \sum_{n=0}^{\infty} \frac{1}{\left(2n+1\right)^2} e^{-\frac{\kappa (2n+1)^2 \pi^2 t}{L^2}} \right) \qquad (23)$$

In comparing the bathymetry results of VBA model with the observational datasets we make use of the same data sets ([57], [26], [48]) and procedure as that employed in the previous study of Stein and Stein [53], allowing thereby direct comparison. The bathymetry data employed refer to values averaged over 2-Ma bins and these have been used also in our comparative analysis. A comparison of VBA model fit to this bathymetry dataset, where the calculations make use of the parameter values given in Table (1), is presented in Figure (8).

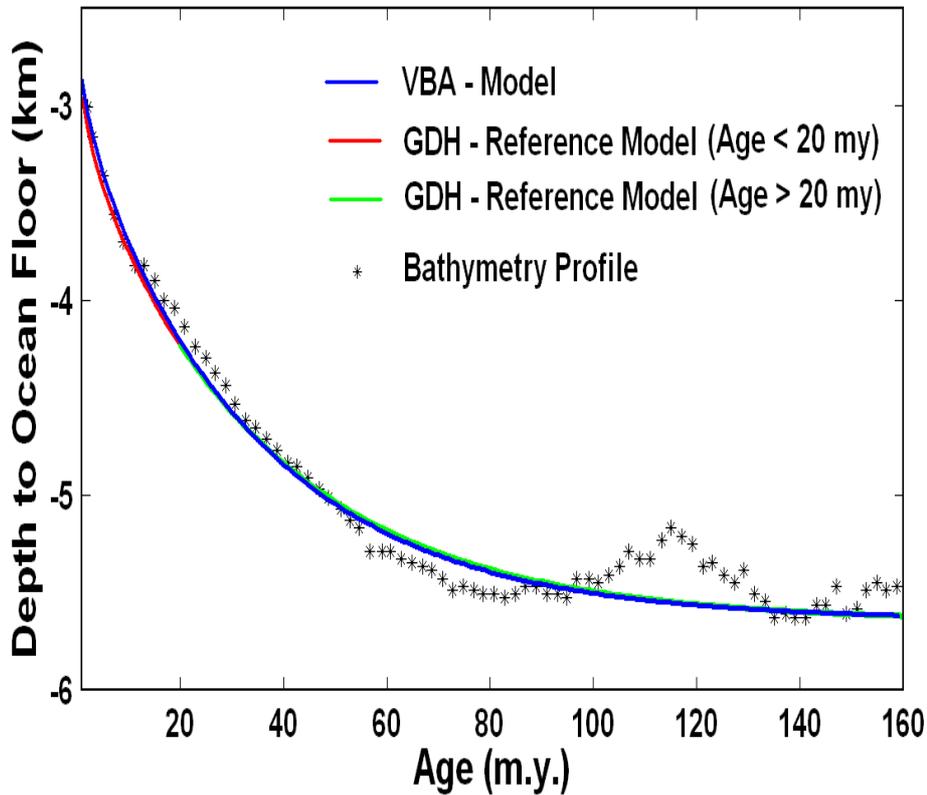

Figure 8: Comparison of the fits of VBA and GDH reference models to bathymetry data. Note that VBA model curve (in blue color) of the present work provides a remarkably good fit for the entire age range of the oceanic lithosphere. The GDH model [53] requires two separate curves (red and green curves), for arbitrarily selected age intervals, in achieving an equivalent fit.

Note that VBA model curve (in blue color) provides a remarkably good fit to bathymetry data (indicated by the asterisk marks) for the entire age range of the oceanic lithosphere. Also shown in this figure are the GDH model curves for bathymetry. The GDH model requires two separate curves (red and green curves), for arbitrarily selected age intervals, in achieving an equivalent fit. Apart from the above mentioned restriction, both VBA and GDH models provide equally good accounts of ocean floor bathymetry. The vertical temperature field of the lithosphere, derived from the VBA model, is similar to the example illustrated in Figure (3). On the other hand, unlike those derived from GDH reference models there are no discontinuities in the temperature field of the lithosphere (Hamza et al. [16]).

At this point it is convenient to consider the sensitivity of VBA model response to the values of the parameters listed in Table (1). For young ocean crust (with ages less than 55 Ma) the main parameter that controls bathymetry is $\delta$, the best fit value of which is 0.6. The dashed and dotted curves in Figure (9) are model curves for $\delta$ values of 0.7 and 0.5 respectively. These model curves bracket the observational bathymetry data for ocean crust with ages less than 55 Ma.

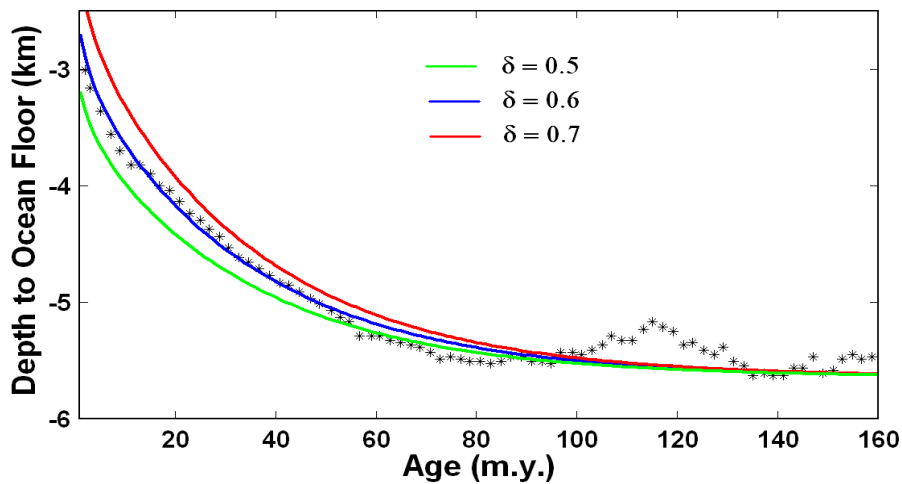

Figure 9: Results of numerical simulations illustrating the response of VBA model to change in the value assumed for the bathymetry constant $\delta$. The square symbols indicate mean oceanic bathymetry values.

For old ocean crust (with ages > 55 Ma) the main parameter that control bathymetry is the basal temperature, the best fit value of which is 1300K. The dashed and dotted lines in Figure (10) are model curves for basal temperatures of 1400K and 1200K respectively. These model curves bracket most of the observational bathymetry data for ocean crust with ages greater than 55 Ma.

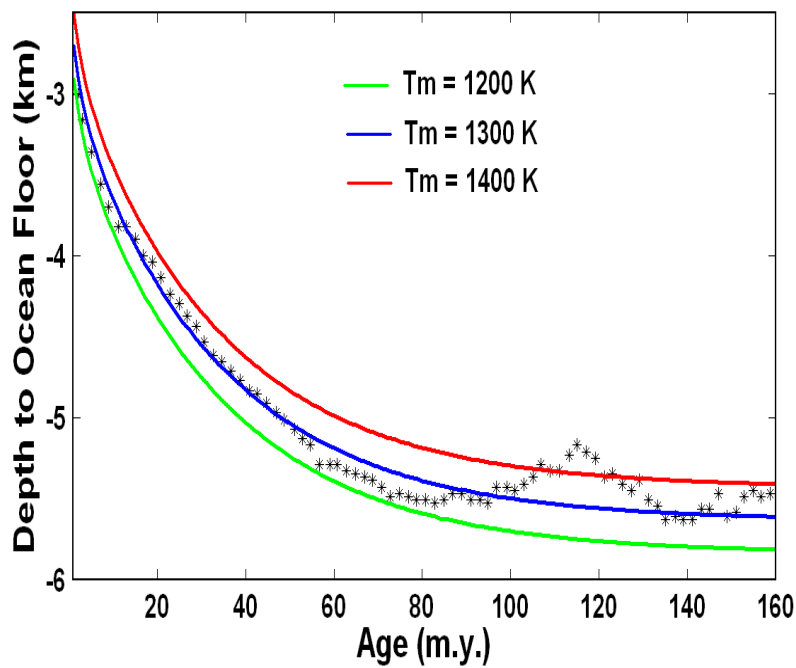

Figure 10: Results of numerical simulations illustrating the response of VBA model results to change in the value assumed for the basal temperatures ($T_i$). The square symbols indicate mean oceanic bathymetry values.

## 5. Discussion and Conclusions

In contrast to HSC and Plate modes, which are closely related regarding the assumptions made and in their implementation, the newly proposed VBA model of oceanic lithosphere assumes that both the thickness and the temperature of the magma rich basal segment vary with distance from the ridge axis. Estimates of regional heat flux in the VBA model are lower than those obtained in previous thermal models of the lithosphere, including the recently proposed Plate model with variable thermal conductivity (McKenzie et al, [31]). Model heat flow values calculated on the basis of equation (9) may be used along with digital isochron data of Muller et al, [33] in mapping heat flow in oceanic regions, following a procedure similar to that suggested recently by Wei and Sandwell, [58]. In addition, theoretical values derived in this manner may be appended with experimental data for the continental regions in deriving global heat flow maps, an example of which is presented in Figure (11). Note that the global map of figure (11) display regional features in heat flow similar to those reported in recent 36 degree harmonic representation of IHFC data set (Hamza et al, [15]). Thus, while ridge areas have heat flow in excess of 80mW/m$^2$, the remaining parts of oceanic regions and continental areas are characterized by heat flow less than 70mW/m$^2$. Also, the global mean heat flow is 61mW/m$^2$ while the maximum binned value is no more than 150mW/m$^2$.

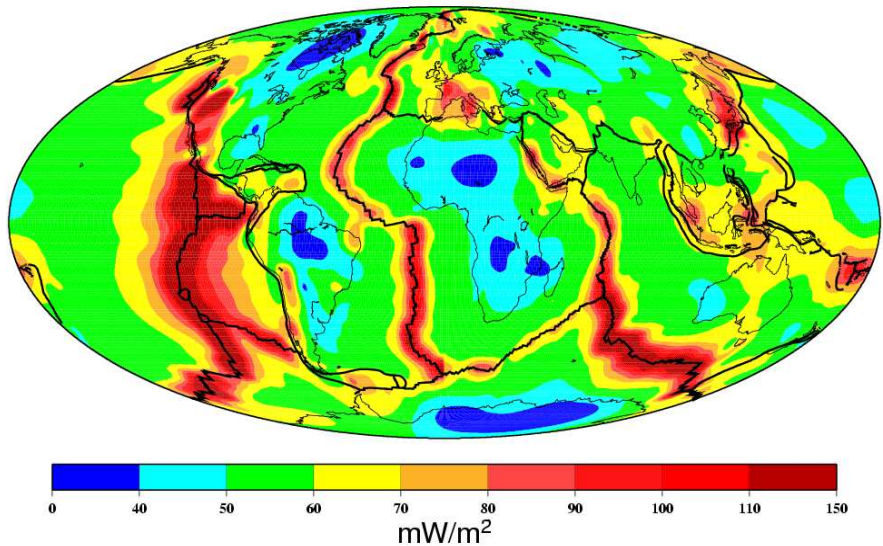

Figure 11: Global heat flow map derived from mixed data sets. For oceanic regions with ages less than 120Ma heat flow values are calculated using equation (14) with age values derived from a global grid at 0.1$^0$ spacing (Muller et al. [33]). For continental regions and remaining areas heat flow values are derived from IHFC data set, following the procedure outlined for higher degree harmonic representation (Hamza et al. [15]).

The discrepancy of previous models of global heat flow from the experimental values has been hypothesized to originate from regional scale hydrothermal circulation in oceanic crust. As mentioned earlier, the validity of the hypothesis of regional scale hydrothermal circulation in oceanic crust is questionable, in view of available information on the thermal and hydrological characteristics of the ocean crust ([20], [16]).

A problem of related interest is the Global Heat Loss. The VBA Model leads to estimates of regional heat flow lower than those derived from previous models, in agreement with recent results of higher degree harmonic representation of global heat flow (Hamza et al, [15]). In the previous study of Pollack et al, [41] the global heat loss was estimated at about 44TW. However, this value is based on the use of mixed data sets, which include both experimental data as well as theoretical values, this latter ones based on the hybrid model of Stein and Stein [53]. Recent work by Hamza et al [15], based on a reappraisal of global heat flow database and with due emphasis on observational data, has concluded that global conductive heat loss falls in the range of 29 to 34TW. This is nearly 23 to 34% less than the previous estimates. We recall that geochemical constraints discussed recently by Hofmeister and Criss, [20] point to the need for downsizing the current estimates of global heat loss.

The main conclusions of the present work may be summarized as:

1 - The new VBA model of the oceanic lithosphere allows incorporation of the thermal effects of variable heat input into its basal parts, whereas previous models (HSC and Plate) take into account only effects produced by surface heat loss;

2 – The width of magma injection zone in the spreading center in VBA model is relatively narrower, and the transition to stable non-magmatic configuration take place on time-scales much shorter than those predicted by the conventional boundary layer theory;

3 – The constant temperature Plate model envisaged by McKenzie [30] is a particular case of the more general class of VBA models;

4 - The VBA Model provides a vastly improved fit to experimental heat flow data for both the younger as well as the older segments of the oceanic crust;

5 – The relation for ocean floor bathymetry derived from VBA model provides an equally good fit to observational data as that provided by the hybrid model curves of Stein and Stein [53];

6 – The VBA model fit to bathymetry data is valid for the entire age range of the oceanic lithosphere. There is no need to introduce ad-hoc adjustments (artificial changes in heat flow) in model fits for ocean floor bathymetry;

7 – The fits of VBA model for sea floor heat flow and bathymetry has been achieved without introducing artificial discontinuities in the temperature field of the lithosphere;

8 – Agreement of the VBA model with the observables exists for any reasonable choice of input parameters. The best agreement is obtained for values closest to those believed representative of the lithosphere, particularly the lowermost extent of the lithosphere;

9 - Estimates of root-mean-square misfit between the VBA Model values and experimental heat flow data are relatively much better than those found for the previous models. Given the uncertainty in marine heat flow measurements and the quality of the fit, there appears to be no need to invoke the hypothesis of regional scale hydrothermal circulation in oceanic crust;

10 - The VBA Model of the present work leads to estimates of regional heat flow that are significantly lower than those derived from previous thermal models of the lithosphere. The new estimates are in reasonable agreement with the results of higher degree harmonic representations of global heat flow (Hamza et al., [15]); and

11 - The current estimates of global heat loss need to be downsized by at least 25%, in support of recent assessments ([20], [15], [16]) and the view that the Earth is in quasi-steady thermal state.

**6. Acknowledgments**

We thank Prof. Anne Hofmeister (Washington University, MO, USA) for critical comments and suggestions which have contributed to significant improvements in the manuscript. The present work was carried out as part of a research project for investigating thermal isostasy in southeast Brazil, with funding from Research Foundation of the State of Rio Janeiro – FAPERJ (Grant No. E-26/100.623/2007), under the program ´Scientist of the State of Rio de Janeiro´. The second author has been recipient of a Ph.D. scholarship granted by Coordenadoria de Aperfeiçoamento de Pessoal de Nível Superior – CAPES, Brazil. The third author contributed to the development of integral transform solution for variable basal heat input, as part of his Ph.D. thesis work on thermal models of continental lithosphere.

Appendix

Thermal model of lithosphere with Variable Magma Accretion in its basal parts

Solution by the Integral Transform Method

In addressing the problem described by equations (3) and (4) of the main body of this paper we introduce the dimensionless variables θ, X, Z, G and $\tau$ :

$$\theta = \frac{T}{T_S}; \; X = \frac{x}{L}; \; Z = \frac{z}{L}; \; G = \frac{L^2 g}{\lambda T_S} \text{ and } \tau = \frac{\lambda t}{\rho C_P L^2} \qquad \text{(A1)}$$

This allows us to rewrite the differential equation (3) for the system of *n* discretized elements as:

$$\frac{\partial \theta}{\partial \tau} + Pe(X) \frac{\partial \theta}{\partial X} = \frac{\partial^2 \theta}{\partial X^2} + \frac{\partial^2 \theta}{\partial Z^2} + G \qquad \begin{matrix} X_j < X < X_{j+1} \\ t > 0 \\ 0 < Z < 1 \end{matrix} \qquad \text{(A2)}$$

where *j* is the discretization index, $X_0 > 0$ and *Pe* is the Peclet number given by:

$$Pe[L(X)] = \frac{v \rho C_P L(X)}{\lambda} \qquad \text{(A3)}$$

The interval $[X_j, X_{j+1}]$ can be chosen to be sufficiently small that the Peclet number may be considered constant within any specific interval. We assume steady state conditions and consider that heat production term *G* is negligible. In this case the differential equation and the boundary conditions become:

$$Pe \frac{\partial \theta}{\partial X} = \frac{\partial^2 \theta}{\partial Z^2} + \frac{\partial^2 \theta}{\partial X^2} \qquad \begin{matrix} X_j < X < X_{j+1} \\ 0 < Z < 1 \end{matrix} \qquad \text{(A4a)}$$

$$\theta(X_j > 0, Z) = \theta(Z) \qquad \text{(A4b)}$$

$$\theta(X, Z = 0) = 1 \qquad \text{(A4c)}$$

$$\theta(X, Z = 1) = 0 \qquad \text{(A4d)}$$

$$\theta(X = 0, Z) = 1 \qquad \text{(A4e)}$$

The purpose of condition (A4b) is to avoid the well known problem of singularity at the position X = 0.

We assume that θ *(X, Z)* may be expressed as:

$$\theta\ (X,\ Z) = u\ (X,\ Z) + k\ (Z) \tag{A5}$$

The problem in *K (Z)* is:

$$\frac{d^2 k}{\partial Z^2} = 0 \qquad 0 < Z < 1 \tag{A6a}$$

$$k(Z = 0) = 1 \tag{A6b}$$

$$k(Z = 1) = 0 \tag{A6c}$$

Hence the solution of problem in *K (Z)* is:

$$k(Z) = (1 - Z) \tag{A6d}$$

The problem in *u (X, Z)* is:

$$Pe\ \frac{\partial u}{\partial X} = \frac{\partial^2 u}{\partial X^2} + \frac{\partial^2 u}{\partial Z^2} \qquad \begin{array}{c} X_j < X < X_{j+1} \\ 0 < Z < 1 \end{array} \tag{A7a}$$

$$u(X = 0,\ Z) = Z \tag{A7b}$$

The condition (A7b) is necessary for the solution u(X, Z) to be compatible with the condition (A4e).

We now admit that the solution of the problem in *u (X, Z)* may be expressed as:

$$u\ (X,\ Z) = \sum_{i=0}^{\infty} \psi_i\ (\mu_i,\ Z)\ .C_i\ (X) \tag{A8}$$

The eigen functions $\psi_i(\mu_i, Z)$ of equation (A8) are associated with the following eigen value problem:

$$\frac{d^2 \psi_i}{dZ^2} + \mu^2\ \psi_i(\mu_i, Z) = 0 \tag{A9a}$$

$$\psi_i(\mu_i, Z = 0) = 0 \tag{a9b}$$

$$\psi_i(\mu_i, Z = 1) = 0 \tag{A9c}$$

The auxiliary problem presented is the eigen value problem, typical of Sturm-Liouville, which has the following properties:

a) the eigen values $\mu_i$ are real, positive and the order of values is such that $\mu_0 < \mu_1 < \mu_2 < \mu_3 ... < \mu_i < \mu_{i+1}$ , where *i = 0, 1, 2, 3 ....*

b) the Eigen functions $\psi_i(\mu_i, X)$ associated with the Eigen values $\mu_i$ are orthogonal.

Solving the auxiliary problem we have:

$$\psi_i(\mu_i, Z) = \sin(\mu_i, Z)$$ (A9d)

$$\mu_i = i\pi$$ (A9e)

The coefficients of the expansion $C_i$ $(X)$, are obtained by multiplying equation (A8) by the operator:

$$\int_0^1 \psi_j(\mu_j, Z) dZ$$ (A10)

and we have: $\displaystyle\int_0^1 u(X, Z)\psi_j(\mu_j, Z) dZ = C_i(X)\sum_{i=1}^{\infty}\int_0^1 \psi_j(\mu_j, Z)\psi_i(\mu_i, Z) dZ$ (A11)

The integral on the RHS of (11) is zero for $i \neq j$ (eigen functions are orthogonal, see Özisik, 1980). For $i = j$, the integral leads to the norm $N_i$, associated with the problem:

$$N_i(\mu_i) = \int_0^1 \psi_i^2(\mu_i, Z) dZ$$ (A12)

Consequently, the unknown $C_i$ $(X)$ is given by:

$$C_i(X) = \frac{1}{N_i}\int_0^1 u(X, Z)\psi_i(\mu_i, Z) dZ$$ (A13)

$$C_i(X) = \bar{u}_i(X) / N_i(\mu_i)$$ (A14)

It is obvious that the transform and its inverse in equation (A14) are:

$$\bar{u}_i(X) = \int_0^1 u(X, Z)\psi_i(\mu_i, Z) dZ$$ (A15a)

$$u(X, Z) = \sum_{i=1}^{\infty}\frac{\bar{u}_i(X)\psi_i(\mu_i, Z)}{N_i(\mu_i)}$$ (A15b)

This procedure transforms the partial differential equation into a system of ordinary differential equations. We adopt the following strategy:

a) Multiply the original equation by the operator: $\displaystyle\int_0^1 \psi_i(\mu_i, Z) . dZ$

$$Pe\int_0^1 \frac{\partial u}{\partial X}\psi_i(\mu_i, Z) dZ = \int_0^1 \frac{\partial^2 u}{\partial Z^2}\psi_i(\mu_i, Z) dZ + \int_0^1 \frac{\partial^2 u}{\partial X^2}\psi_i(\mu_i, Z) dZ$$ (A16)

By the Leibniz rule:

$$Pe\frac{d}{dX}\int_0^1 u(X,Z)\psi_i(\mu_i,Z)dZ=\frac{d^2}{dX^2}\int_0^1 u(X,Z)\psi_i(\mu_i,Z)dZ+\int_0^1\frac{\partial^2 u}{\partial Z^2}\psi_i(\mu_i,Z)\,dZ \qquad (A17)$$

Introducing (A15a) in (A17):

$$Pe\frac{d\overline{u_i}}{dX}=\frac{d^2\overline{u_i}}{dX^2}+\int_0^1\frac{\partial^2 u}{\partial Z^2}\psi_i(\mu_i,Z)dZ \qquad (A18)$$

b) Multiplying the equation of the auxiliary problem by the operator $\int_0^1 u(X,Z)\,dZ$ leads to:

$$\int_0^1 u(X,Z)\frac{d^2\psi_i}{dX^2}\,dZ = -\mu_i^2\int_0^1 u(X,Z)\psi_i(Z)dZ \qquad (A19)$$

Equation (A19) becomes:

$$\int_0^1 u(X,Z)\frac{d^2\psi_i}{dX^2}\,dZ = -\mu_i^2\,\overline{u_i}(X) \qquad (A20)$$

c) Subtracting (A20) from (A18):

$$\frac{d^2\overline{u_i}}{dX^2}-Pe\frac{d\overline{u_i}}{dX}-\mu_i^2\overline{u_i}(X)=\int_0^1\left[u(X,Z)\frac{d^2\psi_i}{dZ^2}-\frac{\partial^2 u}{\partial Z^2}\psi_i(\mu_i,Z)\right]dZ \qquad (A21)$$

d) Developing the integrals on the RHS:

$$\frac{d^2\overline{u_i}}{dX^2}-Pe\frac{d\overline{u_i}}{dX}-\mu_i^2\overline{u_i}(X)=\left[u(X,1)\frac{d\psi_i}{dZ}\bigg|_{Z=1}-u(X,0)\frac{d\psi_i}{dZ}\bigg|_{Z=0}-\psi_i(\mu_i,1)\frac{\partial u}{\partial Z}\bigg|_{Z=1}+\psi_i(\mu_i,0)\frac{\partial u}{\partial Z}\bigg|_{Z=0}\right] \qquad (A22)$$

e) To conclude we use the boundary conditions in (A22) which lead to:

$$\frac{d^2\overline{u_i}}{dX^2}-Pe\frac{d\overline{u_i}}{dX}-\mu_i^2\overline{u_i}(X)=0 \qquad (A23)$$

In obtaining the solution of (A23) it is necessary to make use of the transformed boundary conditions:

$$\overline{u_i}(X=X_n)=\int_0^1 u(X=X_j,Z)\psi_i(Z)dZ \qquad (A24a)$$

$$\overline{u_i}(X=X_j)=\int_0^1 Z\psi_i(Z)dZ \qquad (A24b)$$

$$\overline{u_i}\left(X=X_j\right)=\overline{f_i}$$ (A24c)

where:

$$\overline{f_i}=\int_0^1 Z.\psi_i(Z)dZ$$ (A24d)

The solution of (A23) is:

$$\overline{u_i}(X)=A_i\exp(a_1X)+B_i\exp(a_2X)$$ (A25a)

where:

$$a_1=\left(Pe-\sqrt{Pe^2-4\mu_i^2}\right)/2$$ (A25b)

$$a_2=\left(Pe+\sqrt{Pe^2-4\mu_i^2}\right)/2$$ (A25c)

Obviously, only the solution (A25b) has physical meaning, hence:

$$\overline{f_i}=A_i$$ (A25d)

Consequently the problem in $\overline{u_i}$ is:

$$\overline{u_i}(X)=\overline{f_i}\exp(a_1X)$$ (A26)

and using the inversion formula:

$$u(X,Z)=\sum_{i=1}^{\infty}\frac{\overline{f_i}\exp(a_1X)\ \psi_i(\mu_i,Z)}{N_i(\mu_i)}$$ (A27)

In terms of the dimensionless variables the equations for temperature ($\theta$), temperature gradient ($\partial\theta/\partial z$) and heat flux ($\overline{q}$) may be expressed as:

$$\theta(X,Z)=(1-Z)+\sum_{i=1}^{\infty}\frac{\overline{f_i}\exp(a_1X)\psi_i(\mu_i,Z)}{N_i(\mu_i)}$$ (A28)

$$\frac{\partial\theta(X,Z)}{\partial Z}=(-1)+\sum_{i=1}^{\infty}\frac{\overline{f_i}\exp(a_1X)}{N_i(\mu_i)}\frac{d\psi_i}{dZ}$$ (A29)

$$q(X,Z)=-\lambda\frac{\partial\theta(X,Z)}{\partial Z}=-\lambda\left[(-1)+\sum_{i=1}^{\infty}\frac{\overline{f_i}\exp(a_1X)}{N_i(\mu_i)}\frac{d\psi_i}{dZ}\right]$$ (A30)